\documentclass{article}
\usepackage{arxiv}
\usepackage[utf8]{inputenc} 
\usepackage[T1]{fontenc}    
\usepackage{hyperref}       
\usepackage{url}            
\usepackage{booktabs}       
\usepackage{amsfonts}       
\usepackage{nicefrac}       
\usepackage{lipsum}		
\usepackage{graphicx}
\usepackage{doi}
\usepackage{array}
\usepackage{setspace}
\graphicspath{{images/}}
\usepackage{siunitx}
\usepackage{tabularx}
\usepackage{authblk}
\usepackage{makecell}
\usepackage{float}

\title{Relationships between patenting trends and research activity for green energy technologies}

\author[1]{Regina TUGANOVA}
\author[1]{Anna PERMYAKOVA}
\author[1]{Anna KUZNETSOVA}
\author[1]{Karina RAKHMANOVA}
\author[1]{Natalia MONZUL}
\author[1]{Roman UVAROV}
\author[2]{Elizaveta KOVTUN}
\author[2, 3]{Semen BUDENNYY}

\affil[1]{ITMO University, 
Faculty of Energy and Environmental Technologies, Saint-Petersburg}
\affil[2]{Sber AI Lab, Moscow}
\affil[3]{Artificial Intelligence Research Institute (AIRI), Moscow}

\renewcommand{\shorttitle}{\small\raggedright{Relationships between patenting trends and research activity for green energy technologies}}

\begin{document}
\maketitle

\begin{abstract}
Green technology is viewed as a means of creating a sustainable society and a catalyst for sustainable development by the global community. It is responsible for both the potential reduction of production waste and the reduction of carbon footprint and CO$_2$ emissions. However, alongside with the growing popularity of green technologies, there is an emerging skepticism about their contribution to solving environmental challenges.  
This article focuses on three areas of eco-innovation in green technology: renewable energy, hydrogen power, and decarbonization. Our main goal is to analyze the relationship between publication activity and the number of patented research results, thus shedding light on the real-world applicability of scientific outcomes. We used several bibliometric methods for analyzing global publication and patent activity, applied to the Scopus citation database and the European Patent Office's patent database. 
Our results show that the advancement of research in all three areas of eco-innovation does not automatically lead to the increase in the number of patents. We offer possible reasons for such dependency based on the observations of the worldwide tendencies in green innovation sphere. 



\end{abstract}


\section{Introduction}

The global focus of humanity on sustainable development to ensure a stable life and reduce environmental problems requires businesses and the research sector to apply and develop more resource and energy-saving technologies (\cite{fernando2017impact}). Providing businesses with green technologies is a critical part of their activities. Through the implementation of environmentally friendly and efficient developments, a significant reduction in the carbon footprint is possible (\cite{dou2021does}). A crucial  contribution to the achievement of sustainability is made by technological innovations in the field of green technologies, which are a key factor in promoting sustainable development (\cite{Shen_2021, silvestre2019innovations}). Moreover, the observance by technological innovations of production-saving and resource-saving approaches will make it possible to achieve the most coordinated development and the simultaneous solution of existing economic, social and environmental problems (\cite{song2019technological}). 

From the point of view of eco-innovation, it is assumed that focusing on green technologies will bring novelty and innovative change to the functioning of any industry in the same way as information technology (\cite{iravani2017advantages}).

An important role in the development and implementation of green technologies is played by environmental, social and governance (ESG) performance indicators of companies. Despite the influence of ESG in the field of business and technology, a confrontation between scientists on the world stage is indicated. One side supports ESG ratings, declaring their effectiveness at the expense of reputation and competitiveness (\cite{buallay2018sustainability,cappucci2018esg}). At the same time, another part of the researchers believes that ESG ratings are ineffective, since it is not possible to evaluate the true value of criteria, but it is possible to fulfill all the conditions and requirements of the ratings without additional efforts (\cite{garvey2016pitfall,avetisyan2017consolidation,entine2003myth}). This trend is observed due to the fact that the mechanism of the relationship between ESG ratings and green innovations has not been studied in detail, therefore, the results do not act as a universal criterion for evaluating green technologies (\cite{chouaibi2021esg}). Existing research is also limited and, in most cases, based on developed countries with relatively mature ESG systems, as developing countries are not well established on the market and are still in the early stages of creating an ESG system (\cite{tan2022effect}).  Although ESG ratings have a significant impact on ecoinnovation in the field of green technologies, these energy-efficient and resource-efficient processes play a predominant role in the development of the industrial sector.

Interestingly, in recent years, the existence of an industrial paradox can be observed with the growing demand for renewable energy technologies (RES). Apparently, the ESG agenda in part of the letter E catalyzes interest in carbon-neutral energy, and RES is the brightest representative of this energy. In other words, the ESG agenda creates additional demand for renewable energy, catalyzes this market, and promotes more investment. However, the production of renewable energy will require the more forced extraction of rare-earth metals, such as nickel and lithium. In addition to other metals often used in the production of electric batteries, they are by far the main capacity for RES-generated electricity. Consequently, increasing the rate of extraction of target metals for RES will increase emissions, increasing the anthropogenic impact on the environment. Despite the initial good intentions to improve the environmental situation (\cite{article_lebre}).

Over the past decades, the industrial sector has faced two main constraints: the waste of natural resources and the rapidly growing pollution of the environment (\cite{cai2020can}). Therefore, ecoinnovations have become an important measure to solve the problems of developing industries that negatively affect the state of the biosphere. Consequently, production is indispensable in the concept of sustainable development, whose vector aims at the development of new environmentally friendly technologies. One such example is the use of environmentally friendly chemicals from renewable raw materials that contribute to the sustainability of the chemical sector. However, these “green” chemical technologies can only contribute to a more sustainable society if their environmental impact is lower than that of their traditional counterparts (\cite{thomassen2019assess}).

For the development of the industrial technology sector, new ideas, methods for implementing, developing and testing prototypes of resource- and energy-efficient green technologies in smaller laboratory conditions using experimental facilities are needed. To implement this research trajectory, research and development (R\&D) is widely used around the world.

Sponsoring R\&D acts as a contribution to innovation. According to the R\&D  specifics, not all the results of this work correspond to technological advances, which are usually perceived as the result of innovative activity. Some countries, such as China, allocate government subsidies to stimulate the introduction of R\&D  in the field of green technologies, which does not always contribute to their development and subsequent implementation on a large scale (\cite{hu2019coupling}). It is worth noting that in recent years there have been a growing number of studies, the object of which is the quantitative indicator of patent activity in the field of environmental innovations in green technologies. Patents are considered the result of research experimental activity; however, not all patents can have practical application in production. Therefore, researchers try to measure technological progress based on the production structure (\cite{du2019towards}).

Despite numerous theoretical studies showing a direct proportional relationship between the number of innovative ecotechnologies and the improvement of the climate situation, there is weak empirical evidence to confirm this (\cite{song2019technological}). The ambiguity of the impact of eco-innovations in the field of green technologies on reducing the carbon footprint was also noted. This was expressed in the presence of both positive and negative dynamics of the impact of eco-innovations on CO$_2$ emissions, depending on the conditions. The varied impact can be observed due to several factors: the level of income of the country and the time interval (\cite{du2019green}), which allows us to conclude that the contribution of ecoinnovations to sustainable development is ambiguous.



The field of green technology affects all areas of human activity (\cite{du2019towards,wang2019green,ikram2021assessing}). Some of the most promising tracks, namely renewable energy technology, decarbonization, and hydrogen energy, were chosen for the study in this article (\cite{oliveira2021green,thomas2020decarbonising,falcone2021hydrogen}). It is these tracks contribute to the dynamic development of the economy and science in general. Furthermore, selected areas are actively supported by the Sustainable Development Goals (SDGs) (\cite{gliedt2018innovation}). These are 'Affordable and Clean Energy' (SDG number 7) (\cite{hak2016sustainable}), 'Industry, 'Innovation and Infrastructure' (number 9) (\cite{leal2021industry}), 'Responsible Consumption and Production' (number 12) (\cite{wang2019literature,bengtsson2018transforming}), and 'Climate Action' (number 13) (\cite{du2019green}).

In recent times, Artificial Intelligence (AI) has been introduced in many spheres of people's life. Such wide application is entailed with intensive training of machine learning models and, as a consequence, with significant energy consumption. Hence, there is a serious concern about direct connection between AI computing and considerable CO$_2$ emissions. A special tracker \cite{https://doi.org/10.48550/arxiv.2208.00406} can be leveraged to track carbon emissions resulting from the training procedure. 

The purpose of this work is to analyze the connection between publication and patent activity in the field of renewable energy, decarbonization, and hydrogen energy. 

\section{Literature review}

Over the past few centuries, research into non-renewable natural resources and the production of electricity from them has been of great importance in meeting all human needs and society as a whole. According to studies, it has been found that the worldwide demand for electricity is only increasing every year, and compared to the level of 2010, by 2040 the demand for energy will increase by 56\% (\cite{Rahman}). However, the production of necessary energy in power plants is accompanied by the active production of greenhouse gases such as carbon dioxide and carbon monoxide.  The formation of such emissions in excessive amounts is the main cause of global warming, and worldwide dissatisfaction with the negative impact of the energy sector on the environment has increased in recent years. 

Approximately 35\% greenhouse gases are emitted by existing power plants, making the energy sector one of the main contributors to global warming (\cite{maamoun2020identifying}). For this reason, alternatives to generate energy in an environmentally friendly way have been actively developed in recent decades. Therefore, the worldwide implementation of renewable energy sources (RES) in the energy sector is intended to improve the biosphere and prevent the possible effects of global warming, which leads to the sustainable development of society (\cite{NYASAPOH2022e01199}).

Renewable energy is generated from inexhaustible natural resources, so this type of energy significantly reduces society's dependence on traditional energy sources such as oil, natural gas, and coal. The application of RES in different areas of life is one of the most promising and environmentally friendly solutions to combat the reduction of greenhouse gases and to contribute significantly to the decarbonization of the energy sector (\cite{bookBarwant,sharif2021role,rahman2022environmental}). However, one cannot ignore the fact that many scientific articles question only the positive effects of RES, so studies actively discuss the negative impact on the environment of several specific RES (\cite{rodrigues2018estimation}). 

Figure 1 shows the distribution of the capacity generated from each type of renewable energy over the period from 2020 to 2022. According to the results, hydropower takes the lead in terms of power generation over the period of time presented. Wind, solar photovoltaics and bioenergy produce less power each year. In 2021, wind power was 1,818.5 TWh, solar photovoltaics was 994 TWh, and finally bioenergy generated 667.2 TWh. The contribution of geothermal, concentrated solar, and ocean energy is a very small percentage, but the three types of energy have the potential for further development and active implementation.

\begin{figure}[h!]
\begin{center}
\includegraphics[scale=0.6]{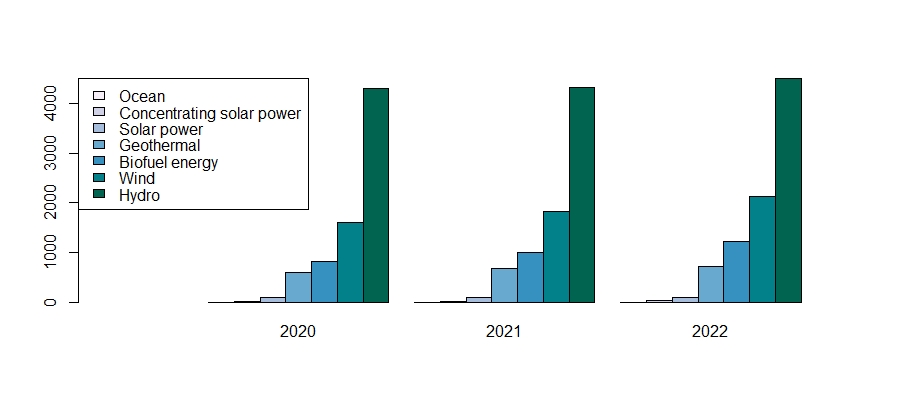}
\caption{Annual capacity generated by renewable energy sources, 2020-2022 (TWh)}
\label{figure1}   
\end{center}
\end{figure}

Studies show that increasing the amount of energy generated through renewable energy sources has a positive effect on reducing carbon dioxide emissions only if the renewable energy consumption threshold is overcome (\cite{CHEN2022117}).

Despite the introduction of renewable energy processes, the main type of energy is derived from fossil fuels, especially petroleum. Yet oil is a dominant source of energy, it still faces the problem of carbon intensity because the world is not ready to abandon energy derived from fossil fuels (\cite{griffiths2022decarbonizing}). 

As a method of reducing CO$_2$ emissions, in addition to the introduction of renewable energy technologies, decarbonization strategies are considered. One of the main methods of decarbonization is the strategy of increasing operational efficiency through the implementation of energy-saving processes. In addition, energy conservation in some cases allows for a significant reduction in the cost of production. In metallurgy, for example, energy costs can reach 40\% of the cost of production. The use of traditional energy sources, such as coal, leads to the fact that metallurgy is responsible for 7-9\% of global carbon dioxide emissions(\cite{kim2022decarbonizing}).

Most oil and gas companies continue to focus on improving operational efficiency and reducing costs by identifying and implementing best practices, continuous improvement programs, and strict compliance monitoring (\cite{griffiths2022decarbonizing}). With increasing demands for environmentally friendly products and reductions in carbon dioxide, the solution for many heavy industry companies is methods and concepts to improve the operational efficiency of processes. An example of such a concept is Green Lean Six Sigma (GLSS), which allows you to reduce the negative impact on the environment, reducing waste and emissions, while improving the company (\cite{mohan2022green}). 
 
The principles of a circular carbon economy are also used to reduce carbon dioxide emissions and decarbonize. It aims to reduce resource consumption while maintaining the output of goods and services and involves reuse and recycling, as well as recycling that can not be reused. This model of behavior allows for a transition to a more sustainable use of resources, minimizing the amount of resources and waste involved, and, as a consequence, greenhouse gas emissions. The basic principles of this approach are the “4R”: reduce, reuse, recycle, remove (\cite{yu2021environmental}). 

Most decarbonization methods can be classified as “reduction” methods, as currently the main method of decarbonization is the overall reduction of carbon dioxide emissions (\cite{craig2018carbon,bistline2021impact}). At the same time, there are very few technologies that combine several R methods, but integrated approaches are necessary for industries such as the oil and gas and metallurgical industries (\cite{griffiths2022decarbonizing}). 

Since oil production requires extensive drilling and involves huge volumes of drill cuttings, there is great potential to reduce the carbon footprint by changing standards for drill cuttings disposal, in particular cuttings re-injection technologies to enhance oil recovery (\cite{xiong2022study}).

Disposal of released unused thermal energy using new technologies is becoming a promising topic. In accordance with traditional practice, generating plants are used for this purpose, which operate on the basis of the organic Rankine cycle technology (\cite{loni2021review,loni2021critical}). Another option is cogeneration, which allows us to use the heat generated in electricity generation in production, oil refining, and chemical processing operations (\cite{celebi2019next}).

Reuse refers to methods to use CO$_2$ that has been captured (\cite{mikulvcic2019flexible,kim2018integrated}). In addition, there are developed methods to use CO$_2$ for beverage and food production, refrigeration, firefighting, water treatment and health care, fertilizer production, and enhanced oil recovery. Consequently, well scaled technologies for synthetic fuels, chemicals, and the use of CO$_2$ in the construction industry are being developed (\cite{schemme2020h2,hepburn2019technological}). The reuse of CO$_2$ can play an important role in the circular carbon economy, turning waste into valuable products and reducing the carbon footprint.


Hydrogen is considered to be a promising option for the future energy carrier of the energy industry, and its study has been conducted actively for many years. This is justified by the fact that this compound has the potential to generate heat and electricity, provide energy to vehicles, as well as energy storage and storage systems with minimal or no environmental impact at any scale and methods of its application. Also, hydrogen has such an advantage as the absence of carbon dioxide emissions when it is used as an energy carrier, since as a result of the combustion of hydrogen fuel, water is formed, which again enters a closed hydrogen production cycle (\cite{Pareek2020InsightsIR}). However, any transition from a traditional energy system based on carbon and fossil fuels to an economy based on hydrogen energy is inextricably linked and hampered by significant scientific, technological, and socioeconomic barriers (\cite{Edwards2007HydrogenE}).

According to the physical parameters, the low density of hydrogen is the main critical factor to ensure stable storage and transportation of hydrogen and, therefore, to facilitate the transition to an economy focused on hydrogen energy (\cite{Abe2019HydrogenEE}). It is also characterized by high costs for the use of hydrogen at the moment. 

At the moment, according to recent studies, there are three main types of hydrogen storage systems: gaseous, liquid, and solid phase storage systems, which differ depending on the size of the storage and the area of hydrogen use (\cite{Abe2019HydrogenEE, niaz2015hydrogen,zhang2016survey}).

The most established storage system for hydrogen gas is physical storage as pressurized hydrogen, but the disadvantage of this type of storage is the need to maintain high pressure and very low temperatures. With this method, there is also a high probability of hydrogen leakage (\cite{sherif2014handbook}). All of the above is the main reason for the incorporation of this method of hydrogen storage into large-scale applications. The next type of hydrogen storage is cryogenic liquid storage. This technology requires maintaining low temperatures, which consume up to 30\% of the total energy (\cite{midilli2005hydrogen}).This is what limits the application of this method of hydrogen storage. A promising variant of hydrogen storage is its binding to solid carriers, such as metal hydrides. However, none of the presented metal hydrides acts as a stable storage method, since questions about low storage capacity of hydrogen, the kinetics, and temperatures of hydrogen absorption and adsorption are still open (\cite{Abe2019HydrogenEE, eftekhari2017electrochemical}). Thus, the types of hydrogen storage studied are not perfect and stable enough to use these technologies on a scale, and there are no norms or standards for hydrogen storage systems.

There are also significant limitations and barriers to the implementation of hydrogen transportation. The main problems lie in the infrastructure itself for the distribution of hydrogen to the consumer, which is currently carried out by land and sea transport. In addition, the barriers to the use of hydrogen energy include high costs associated with the loss and inefficient use of energy due to the low volume density of hydrogen during transportation (\cite{abdalla2018hydrogen}). Consequently, the remaining barriers to the integration of hydrogen into the global economy are its storage and transportation.

\section{Methodology}

A methodology for evaluating publication and patent activity was developed to investigate their correlation. It consists of three major steps:

\begin{enumerate}
    \item \textbf{Key words accumulation} based on their frequency of occurrence in green technology research: renewable energy, decarbonization, and hydrogen energy. The most frequently used words are examined further.
    
    \item \textbf{Examination of publication activity}. It was carried out by comparing the number of publications on a specific topic over an annual period, using some bibliometric methods that have been widely used to identify current trends in various research fields (\cite{Liu2019,article}). The Scopus database was chosen for this research, as it is the leading international bibliographic platform that provides a multidisciplinary citation data on peer-reviewed scientific publications (\cite{Zahra2021}). The number of publications was obtained from the database after applying specific search filters such as publications only in English, article format, the presence of specific keywords, and the publication of the article in the specified year. The use of only one database and the filters listed above impose certain geographic and language coverage limitations; however, in this case, they do not violate the representativeness of the processed data for the study. As the calculation period, the interval 2011-2021 was chosen because it allows for the inclusion of the most recent actual developments and research in the analysis. Publications from 2022 are excluded from the statistics due to the distortion of the dynamics caused by incomplete data. Mendeley, an online bibliographic manager, was used to examine decarbonization publication activity from 2014 to 2021 (\cite{elston2019mendeley,katchanov2019comparing}).
    
    \item \textbf{Analysis of patent activity}. Patents were searched using the same keywords that were used to analyze publication activity. The data for the European Union was obtained from the European Patent Office database, and each keyword was searched for each year. The European Patent Office was used for patent research because it is the executive body of the European Patent Organization and decides on the granting of patents for inventions. As a result, the collected numerical data on the number of patents was analyzed graphically for the last ten years, starting from 2011. 
    
\end{enumerate}

A comparison of publication and patent activity using frequency diagrams makes it possible to determine the ratio of the number of studies and their direct practical result in the form of the result of intellectual property.

\subsection{Data collection for subsequent analysis}

Bibliometric analysis includes mathematical and statistical methods (\cite{donthu2021conduct, Daim2006}). The applied method allows to evaluate the actual popularity of the topic.
Citation bases (Scopus, Web of Science) serve as a data source for bibliometric analysis. In this article, the analysis was carried out based on the Scopus database, as suggested in some studies under consideration (\cite{nobanee2021bibliometric, tamala2022bibliometric}).
The first step in the application of bibliometric analysis is to accurately determine the number of studies to be analyzed. Failure to follow this step may result in errors and incorrect data. It is equally important to create a sample sufficient for analysis, so the number of references should exceed 200 (\cite{tamala2022bibliometric}).
For the analysis, samples were collected from the Scopus database, created using keywords for each area of green energy considered. Keywords were selected based on their most frequent use in articles on the topic of green energy, decarbonization, hydrogen energy, and renewable energy. The keywords were found in the TITLE-ABS-KEY fields. 

Due to the large number of articles in the research areas under consideration, it was decided to limit the sample to publications published in 2021 only. The number of articles that matched the request, on the other hand, exceeded several thousand. Taking into account the Scopus database's export restrictions, three CSV files with a total of 2000 articles were received.

\section{Results}

\subsection{Publication activity}

A review and analysis of publication activity in the Scopus citation database was performed to identify general trends in the development of research in the field of renewable energy and hydrogen energy. The Mendeley system was used to analyze the decarbonization publication activity, and the same search filtering method was used. The analysis reveals a general upward trend in the number of publications. A comparison of publication and patent activity was also used to identify the direct implementation of new technologies and research results in the real world. 

The most frequently encountered keywords in the studies on the topic, identified during the preliminary analysis, were used as keywords. Types of RES were used as keywords, such as: carbon-neutral fuel, biofuel energy, geothermal energy, wind energy, solar energy, and water energy. The results of the publication activity over the 10-year period by keyword are shown in Figure 2a. Keywords for decarbonization: decarbonization, energy sector, carbon-neutral, CO$_2$ emissions, carbon capture and storage, climate mitigation. Common keywords were used to compile the figure 2c, specifying the direction of the main weaknesses of hydrogen technologies, storage and transportation.

\begin{figure}[h!]
\includegraphics[scale=0.46]{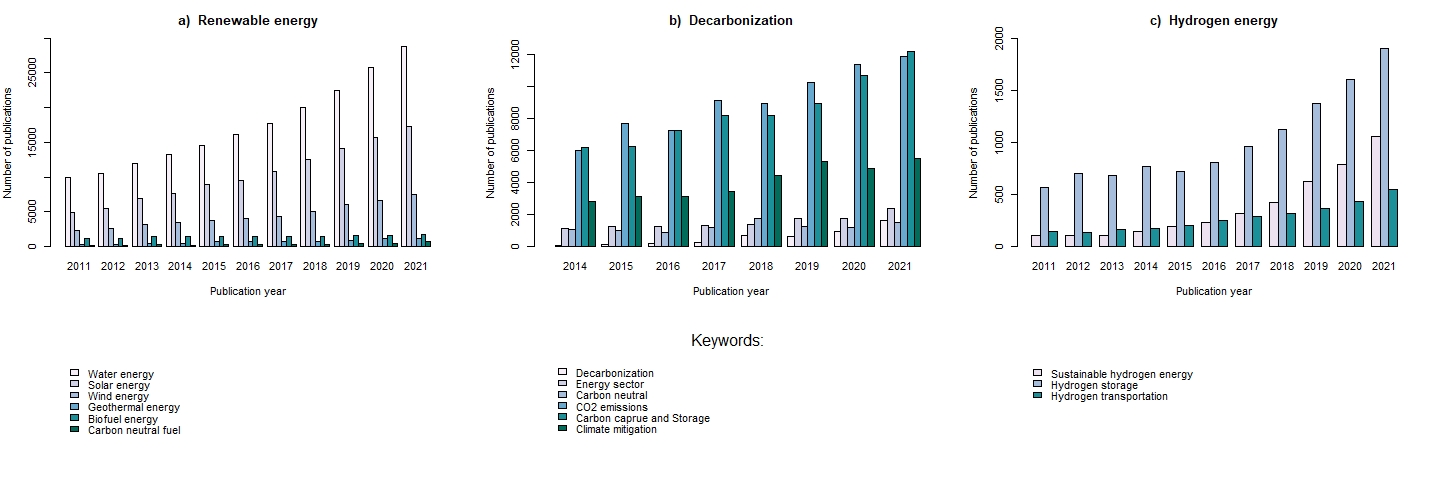}
\caption{Publication activity for renewable energy, decarbonization, hydrogen energy during the past 10 years according to the Scopus database}
\label{figure1}   
\end{figure}

Figure 2a shows a significant increase in the number of publications on specific types of RES, namely solar and hydropower. The availability of such technologies, as well as their potential for widespread dissemination, may determine the persistence of scientific interest in the development and improvement of the efficiency of such processes. Wind energy has seen a slightly more moderate linear increase in the number of publications. It is also worth noting that between 2011 and 2015, there was a surge in scientific interest in geothermal energy, which led to an increase in publications. Furthermore, the field of bioenergy and carbon-neutral fuels has remained stagnant due to a lack of technological advancement or a decline in publication activity. Therefore, it can be assumed that these areas of RES are not significantly popular in research papers in recent years.

Figure 2b depicts a general trend of increasing decarbonization publication activity from 2014 to 2021. Simultaneously, the topic of decarbonization itself becomes significantly more prominent in 2020. The issue of carbon dioxide emissions, which is directly related to decarbonization processes, is of great interest.

According to Figure 2c, the field of hydrogen storage shows unstable growth until 2014, when it begins to decline in terms of publication activity. However, since 2016, there has been a sharp steady increase in the number of scientific publications worldwide in the field of hydrogen storage; by 2021, the rate of publication activity worldwide has increased fourfold compared to 2011. However, the publication activity index in the field of sustainable hydrogen energy has shown a consistent exponential growth from 2011 to 2021. The number of scientific papers on hydrogen transportation is steadily increasing.

According to data obtained in the last five years, the field of renewable energy sources has the highest number of publications (57\%), the second place is decarbonization (28.7\%), and the least interest is in hydrogen energy (14.3\%).

\subsection{Patent activity} 

Similarly to the publication activity, the European Patent Office database was used to analyze registered patents in the fields of renewable energy sources, decarbonization, and hydrogen energy. The frequency diagrams created using the data allow for the tracking of certain trends in the application of new technologies and their market introduction in three specific areas (Figure 3).

\begin{figure}[h!]
\includegraphics[scale=0.46]{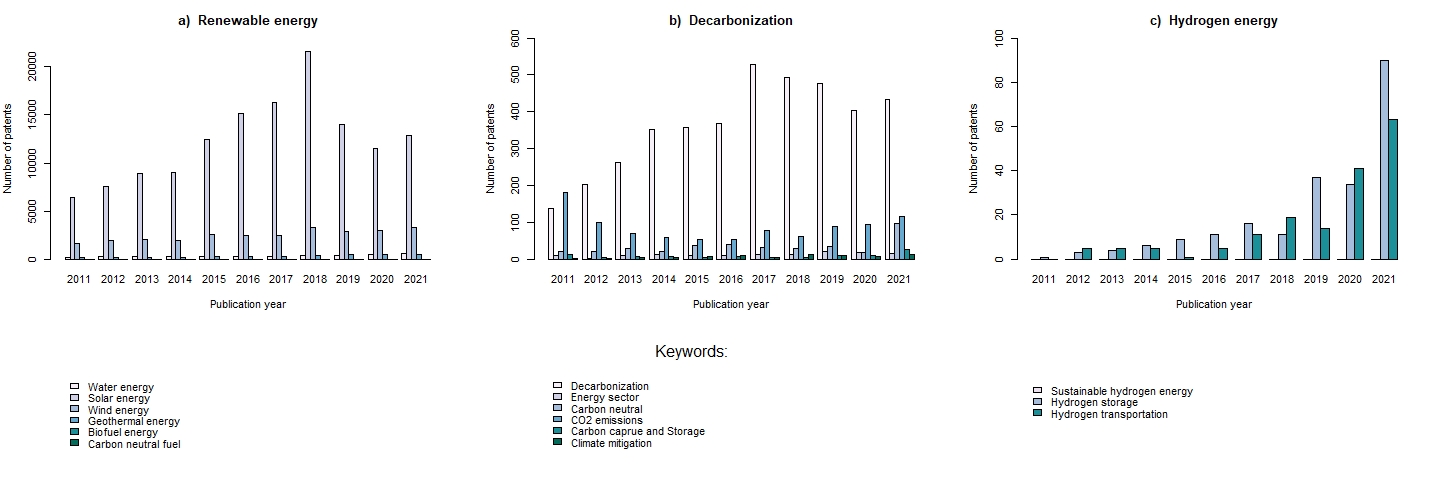}
\caption{Patent activity in the past 10 years according to the European Patent Office}
\label{figure1}   
\end{figure}

According to RES data, a significant number of publications account for hydrogen power direction, thus determining the overall dynamics in all RES tracks. Figure 3a depicts the increase in the index of RES patent activity until 2018, whereas there is no dynamic change in this indicator in 2013 and 2014. However, following the peak in 2018, there is a sharp decline in R\&D  activities from 2019 to 2021.

Figure 3b depicts an unstable increase in patent activity in terms of decarbonization. Decarbonization is the most important contributor to the overall dynamics of these studies. In turn, the indicator of publication activity in the field of CO$_2$ emissions is marked by a decrease in the number of patents from 2012 to 2018, followed by an increase until 2021.

Figure 3c shows the overall growth rate of research in the field of hydrogen energy from 2011 to 2021. The indicator of publication activity for the general keyword 'hydrogen energy' is increasing exponentially in this field of study. When looking at "hydrogen storage," there is only a slight upward trend from 2018 to 2021. However, there is an increase in the number of patents registered in the field of hydrogen transportation from 2019 to 2021.

According to the obtained data for the entire study period from 2011 to 2021, the greatest number of patents is observed in the field of renewable energy sources (95.4\%), the second place is decarbonization (2.9\%), and the least interest is in hydrogen energy (1.7\%).

\section{Discussion}

The method of publication and patent analysis was used to investigate the link between scientific and patenting tendencies in the field of green technology. For this purpose, keyword activity in the various databases of scientific publications and patents was evaluated. According to the study, there is a positive trend in global publication activity in all three areas of green energy, namely a stable linear growth for all indicators in all areas. The analysis of patent activity reveals a disparity in the volume of registered patents between the selected green technology tracks. More specifically, over 95\% of patents account for scientific implementations in the field of renewable energy sources. Furthermore, only in the field of hydrogen energy is a stable accelerated dynamics of patent activity growth revealed. This indicator is declining in the fields of renewable energy sources and decarbonization.

As a result, the increased publication activity reflects researchers' increased attention and interest in the investigated areas of green technologies, namely renewable energy sources, decarbonization, and hydrogen energy. The patent activity in these areas exemplifies the situation, implying the difficulty of implementing and practicalizing technological research in green technologies. As a result, current research is difficult and weakly led to the creation of intellectual property results and their subsequent implementation in manufacturing processes and real life.

\section{Conclusions}

According to the results of the analysis of scientific articles on three areas of green technologies – renewable energy sources, decarbonization and hydrogen energy, the relevance of these studies was confirmed. Interest in these areas was noticed not only in the scientific and technological sector but also in the global economy. For example, many states contribute to the development of green technologies by funding research and development (R\&D ). Furthermore, in turn, the global business sector, involved in green technology, emphasizes the ESG agenda. 


At the moment, green technologies are a popular and actively discussed scientific topic in the global community. But despite this trend, we can observe the opposite situation in the practical field of application, which in general contradicts the standard ways of development of research topics into practically applicable results.

\section*{Competing interests}
The authors declare no competing interests.

\bibliographystyle{unsrt}
\bibliography{biblio}

\end{document}